\documentclass[aps,preprint]{revtex4}
\usepackage{amsmath,amssymb,amsthm,mathrsfs,amsfonts,dsfont} 
\usepackage{graphicx}
\usepackage{subfigure}
\usepackage{amsbsy}
\usepackage{bm}
\usepackage{xcolor}
\usepackage{dcolumn}
\newcommand{\blue}[1]{{\color{black}#1}}
\usepackage{algorithm}
\usepackage{algpseudocode}
\usepackage{amsmath}
\usepackage{array}
\usepackage{booktabs}
\usepackage{physics}

\def\bx{\mathbf{x}}

\def\J{\mathbf{J}}

\begin{document}

\title{Response function as a quantitative measure of consciousness in brain dynamics}
\author{Wenkang Du$^{1}$}
\author{Haiping Huang$^{1,2}$}
\email{huanghp7@mail.sysu.edu.cn}
\affiliation{$^{1}$PMI Lab, School of Physics,
Sun Yat-sen University, Guangzhou 510275, People's Republic of China}
\affiliation{$^{2}$Guangdong Provincial Key Laboratory of Magnetoelectric Physics and Devices,
Sun Yat-sen University, Guangzhou 510275, People's Republic of China}
\date{\today}

\begin{abstract}
	 Understanding the neural correlates of consciousness remains a central challenge in neuroscience. In this study, we investigate the relationship between consciousness and neural responsiveness by analyzing intracranial ECoG recordings from non-human primates across three distinct states: wakefulness, anesthesia, and recovery. Using a nonequilibrium recurrent neural network (RNN) model, we fit state-dependent cortical dynamics to extract the neural response function as a dynamics complexity indicator.
    Our findings demonstrate that the amplitude of the neural response function serves as a robust dynamical indicator of conscious state, consistent with the role of a linear response function in statistical physics. Notably, this aligns with our previous theoretical results showing that the response function in RNNs peaks near the transition between ordered and chaotic regimes---highlighting criticality as a potential principle for sustaining flexible and responsive cortical dynamics.
    Empirically, we find that during wakefulness, neural responsiveness is strong, widely distributed, and consistent with rich nonequilibrium fluctuations. Under anesthesia, response amplitudes are significantly suppressed, and the network dynamics become more chaotic, indicating a loss of dynamical sensitivity. During recovery, the neural response function is elevated, supporting the gradual re-establishment of flexible and responsive activity that parallels the restoration of conscious processing.
    Our work suggests that a robust, brain-state-dependent neural response function may be a necessary dynamical condition for consciousness, providing a principled framework for quantifying levels of consciousness in terms of nonequilibrium responsiveness in the brain.
\end{abstract}

 \maketitle

\section{Introduction}
Understanding the neural basis of consciousness remains one of the most challenging questions in neuroscience~\cite{Neuron-2011,TrN-2020,Nature-2025}. 
Despite remarkable advances in neuroimaging and electrophysiology, reliably distinguishing conscious from unconscious brain states purely on the basis of measurable neural activity remains elusive~\cite{PCI-2013,JNS-2015,PCI-2019}. Clinically, this gap limits our ability to monitor depth of anesthesia, diagnose disorders of consciousness, and understand the nature of brain-state transitions among wakefulness, sleep, anesthesia, and coma.
Brain, as a hierarchical system of interacting neurons, lacks a clear theoretical foundation to determine what should be defined and measured in mathematical terms~\cite{Huang-2024}.

A growing number of works suggest that consciousness is not defined by static structural features alone, but by the brain dynamics far from equilibrium~\cite{PNAS-2021,PRE-2021,Galad-2021,PRE-2023,BD-2025}. Empirical evidences now indicate that conscious states are underpinned by rich, flexible, and metastable patterns of neural activity, and thus, dynamics at the edge of instability are essential for maintaining consciousness~\cite{JNS-2015}. For example, dynamical system approaches have revealed that cortical activity during wakefulness is poised near the edge of instability---a critical regime that maximizes responsiveness to external stimuli~\cite{FNC-2014,PNAS-2022}. Complexity analysis showed that attractors in conscious and anesthesia-induced unconscious states exhibit significantly different shapes, which affects the information processing capability~\cite{JNS-2015,Eagleman-2019}.

These empirical observations can be studied within the perspective of nonequilibrium statistical physics. Recent studies show that the human brain, like all living systems, fundamentally operates out of thermodynamic equilibrium~\cite{PNAS-2021,BD-2025}. Measures such as entropy production, broken detailed balance, and probability flux irreversibility have emerged as principled signatures of cognitive complexity and conscious awareness~\cite{PRE-2021,PRE-2023}. For instance, healthy conscious brain dynamics break time-reversal symmetry~\cite{CS-2022}, generating a preferred direction in time (i.e., arrow of time), whereas proximity to equilibrium---as in anesthesia, non-rapid-eye-movement sleep, or neurodegenerative disorders like Alzheimer’s disease---is accompanied by a loss of temporal irreversibility~\cite{Cruzat-2023}.

Our recent theoretical works further support this dynamic complexity by showing that in recurrent neural networks (despite random connections among neurons), the neural response function itself peaks exactly at the edge of chaos, where the system exhibits maximal sensitivity to perturbations~\cite{Qiu-2025,Du-2024}. Within an optimization-based framework for non-equilibrium dynamics, this critical regime naturally emerges as the point where the responsive property of the network is maximized, indicating a deep link between dynamical instability and the network’s capacity for flexible information processing. This is also consistently supported by an analysis of cortical electrodynamics~\cite{PNAS-2022} showing that the information richness occurs at the edge of chaos, and the conscious brain states are located near the edge of chaos. Another study focusing on a clinical index of conscious level (but on recurrent neural networks) also identifies this picture~\cite{Eric-2024}.

In this work, we verify this theoretical concept in real brain dynamics by building a data-driven recurrent neural network, where the couplings between neurons (or brain regions) are learned from a predictive processing of the time series collected 
from experiments~\cite{Data-2013}. 
The recurrent neural network (RNN) model fits high-density electrocorticography (ECoG) recordings 
from the cortex of non-human primates during reversible loss and recovery of consciousness. The main methodology is based on a predictive learning principle~\cite{Yu-2025}.
After the network is reconstructed, we obtain the functional connectivity among all involved brain regions. As derived by our previous theory~\cite{Qiu-2025}, the response function is measured with a tiny current perturbation to the neural dynamics
equation. 
Within the statistical physics framework, 
we provide a principled and mechanistic approach to quantify how cortical dynamics sustain responsiveness far from equilibrium,
and how this capacity deteriorates under anesthesia and re-emerges during recovery. 
While our modeling currently focuses on primate data, the methodology and analysis may lay the foundation for 
future extensions to human datasets and for understanding how the responsiveness of nonequilibrium brain dynamics is correlated with conscious awareness \blue{(such as cortical activity during wakefulness, non-rapid eye movement (NREM) sleep and REM sleep~\cite{Massimini-2005})}.

\section{Model and methodology}\label{model}
In this section, we describe  the network model used to fit the observed time series data. We first give an overall introduction to the training dataset, and then to
the RNN model and training method. Finally, we introduce the response function as a brain complexity index and show how to measure this function from the reconstructed networks.

\begin{figure}[htbp]
  \centering
  \includegraphics[width=0.95\textwidth]{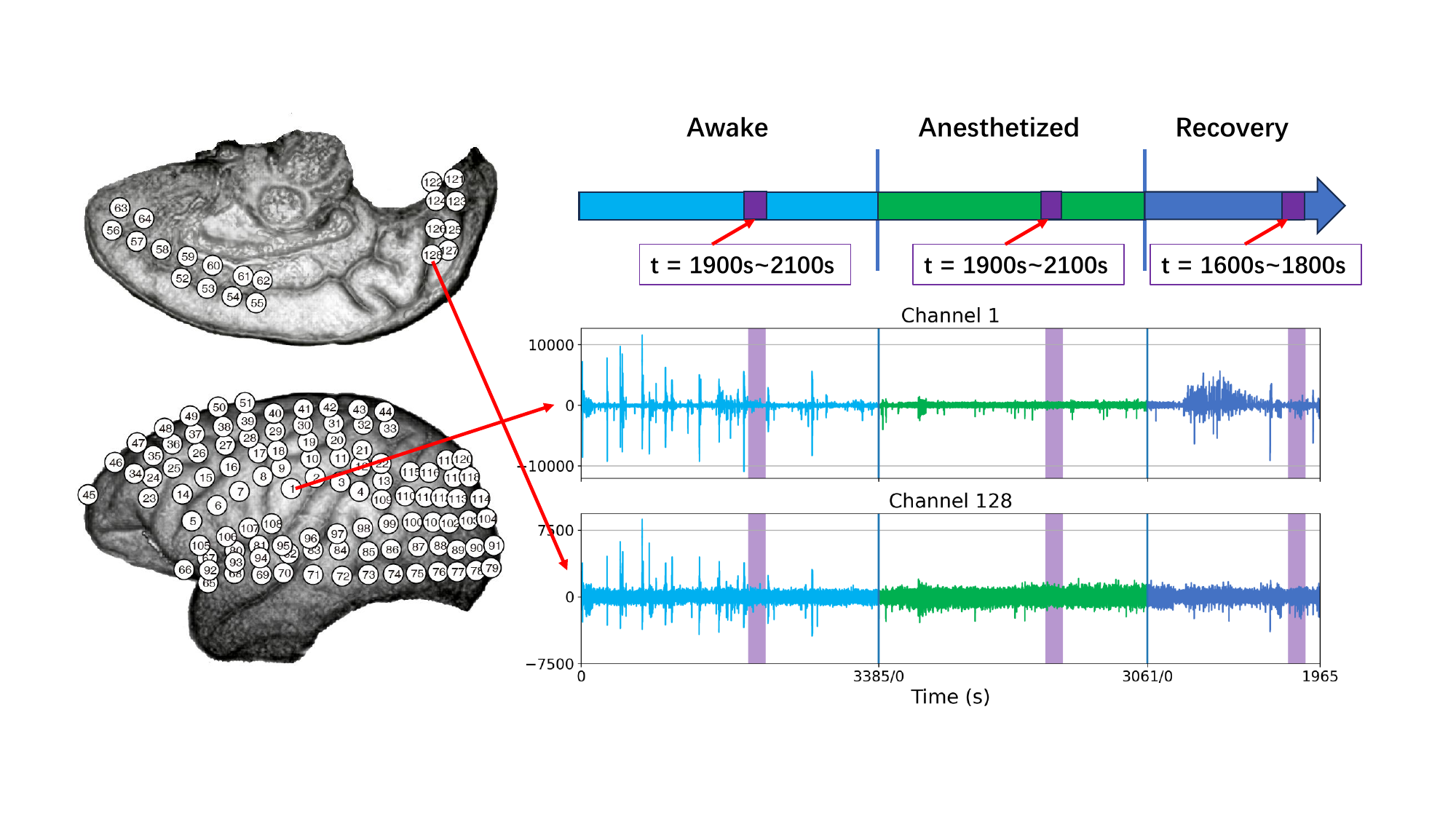}
  \caption{Sketch of experimental data: 2D ECoG electrode array and neural activity signals for Monkey George~\cite{tycho}.
Left: Spatial layout of the $128$‑channel ECoG grid covering frontal, parietal, and temporal cortices 
(image adapted from~\cite{tycho}). 
This map serves as the anatomical reference for interpreting the fitted RNN dynamics across cortical regions. 
Right: Representative neural activity traces from channels~$1$ and~$128$ during three behavioral states---wakefulness, anesthesia, and recovery. 
The purple shaded region marks the time window used for model fitting in this study (up to $200$ s).}
  \label{fig1}
\end{figure}

\subsection{Data description}
In this study, we analyzed electrocorticography (ECoG) recordings from a publicly available dataset of non-human primates provided by the Neurotycho project~\cite{Data-2013,tycho}. 
The data were collected at the Laboratory for Adaptive Intelligence, Brain Science Institute, RIKEN, from four male monkeys under different states of consciousness, including wakefulness, deep sleep, and anesthesia induced by anaesthetic drugs (propofol, ketamine, and ketamine combined with medetomidine). The ECoG array consists of 128 electrodes covering the prefrontal cortice, primary motor and somatosensory cortices, parietal and temporal cortices, and visual cortices, providing high-density cortical coverage. ECoG signals were recorded at a sampling rate of $1 \mathrm{kHz}$. A sketch of this experimental data is given in Fig.~\ref{fig1}.

In the current work, we carried out the analysis of experimental data from a single representative monkey (George), focusing on three distinct recording sessions that capture the transition across conscious wakefulness, induction of anesthesia (loss of consciousness), and recovery (responsive consciousness). While we present detailed results for this example, the same analytical framework and conclusions apply consistently to other settings in the same experiment. A key feature of this dataset is the reversible nature of the anesthetic procedure, allowing us to study transitions across distinct levels of consciousness on the same animal.

The time series data are preprocessed in the following way. We applied a simple standardization procedure by subtracting the mean and dividing
 by the standard deviation for each channel within the time window under consideration.

\subsection{Dynamics model and training}
To capture the dynamical properties of cortical activity in different brain states, we fit the ECoG signal using an RNN model~\cite{Chaos-1988,Yu-2025}. The continuous-time dynamics for each unit $x_i$ are given by
\begin{equation}
\tau \frac{dx_i}{dt} = -x_i + \sum_{j=1}^{N} J_{ij} \, \phi(x_j) + h_i,
\label{eq:rnn_cont}
\end{equation}
where $\tau$ is the neuronal time constant, $J_{ij}$ denotes the recurrent connectivity weight from neuron $j$ to $i$, $\phi(\cdot)$ is a nonlinear activation function ($\tanh$ in this work), and $h_i$ represents a weak 
external sensory perturbation if necessary. In our following analyses, we fix $\tau=1.0$ (but it can be learned as well); in the case of optimizing $\tau$, an additional parameter $\alpha\equiv\Delta t/\tau$ where $\Delta t$ is a small time increment used to discretize Eq.~\eqref{eq:rnn_cont} can be updated to minimize the training cost specified later. We confirm that allowing $\tau$ to be optimized besides $\mathbf{J}$ does not qualitatively change the fitted dynamics or response-function estimates.

To fit ECoG time series, Eq.~(\ref{eq:rnn_cont}) is discretized with a small time increment $\Delta t$ as follows,
\begin{equation}
x_i(t+\Delta t) = (1-\Delta t) \, x_i(t) + \Delta t \sum_{j=1}^{N} J_{ij} \, \phi(x_j(t)),
\label{eq:rnn_discrete}
\end{equation}
where $N=128$ corresponds to the number of ECoG channels, and $\Delta t = 0.001$ corresponds to the sampling interval ($1 \mathrm{kHz}$ in raw data). Note that during fitting, the external perturbation is absent.

To capture the dynamic property of interactions among units, the model is trained on short segments: each training batch contains \blue{$T=2000$ time points (equivalent to two second of ECoG data)}. For each experimental condition (awake, anesthesia, recovery), we repeatedly sample non-overlapping segments from the empirically identified steady-state portions of the recording. These segments are evenly distributed to span over time windows of \blue{$100$ and $200$ seconds} for a robust estimate of the population response.

The network parameters $\mathbf{J}$ are trained to minimize the discrepancy between the observed trajectory $\mathbf{X}$ and the model prediction $\mathbf{\hat{X}}$ using a mean squared error (MSE) loss with an $\ell_2$-norm regularization:
\begin{equation}
\mathcal{L} = \frac{1}{T} \sum_{t} \| \mathbf{\hat{X}}(t) - \mathbf{X}(t) \|^2 
+ \lambda_{\text{reg}} \| \mathbf{J} \|_F^2,
\label{eq:loss}
\end{equation}
where $\| \cdot \|_F$ denotes the Frobenius norm, and $\lambda_{\text{reg}}$ sets the regularization strength (for the current analysis we set $\lambda_{\text{reg}} =10^{-10}$ making the two terms in $\mathcal{L}$ balanced during training). The optimization is performed using the Adam algorithm with a learning rate $0.1$. For each mini-batch, the RNN recursively predicts the neural state at the next time point, and the neural couplings are updated to minimize the reconstruction error~\cite{Yu-2025}.

\subsection{Response function measurement}
To quantify the network's dynamical responsiveness, we estimate the linear response function using a standard perturbative approach, by borrowing concepts from physics~\cite{FD-2008,Qiu-2025}.
 For the neural dynamics, the response function is formally defined as
\begin{equation}
\chi(t) = \frac{\partial}{\partial h} \langle \phi(\mathbf{x}(t)) \rangle,
\label{eq:response_def}
\end{equation}
where $h$ is applied at the zero time point, or the moment when the perturbation is applied, is set to the starting time point.
In practice, $h_i = h$ is applied to all units (a homogeneous perturbation). $\langle\cdot\rangle$ means an ensemble average. Then, the dynamics run after the perturbation according to the following ordinary differential equations:
\begin{equation}
\tau \frac{d\mathbf{x}}{dt} = -\mathbf{x} + \mathbf{J} \, \phi(\mathbf{x}).
\end{equation}
The mean population output $\langle \phi(\mathbf{x}(t)) \rangle$ can then be computed by considering many simulated trajectories.
 The empirical response function is then estimated by the slope of the output-perturbation relation:
\begin{equation}
\chi(t) \approx \frac{ \Delta \langle \phi(\mathbf{x}(t)) \rangle }{ \Delta h }.
\label{eq:response_estimate}
\end{equation}
Once the perturbation is sufficiently weak, as assumed in physics, there exists a linear relationship between evoked response and the perturbation~\cite{Zou-2024}.
In this sense, the intrinsic difference between spontaneous and evoked population activity leads to a quantitative measure for the detection of awareness. 

For each reconstructed functionally connected network, numerical integration is performed using a Runge–Kutta solver with initial conditions set by the recorded ECoG state. 
To ensure robustness, the final response values are computed as averages across multiple steady-state windows ($100$-$200$ seconds) and multiple perturbation amplitudes ($h$ from $0.0001$ to $0.0005$). 
Together, these procedures yield an interpretable estimate of how neural responsiveness varies across brain states, which will be verified in the following, providing a mechanistic link between nonequilibrium cortical dynamics
 and signatures of conscious state, as implied by our previous theoretical work~\cite{Qiu-2025}.

\begin{figure}[ht]
  \centering
  \includegraphics[width=0.95\textwidth]{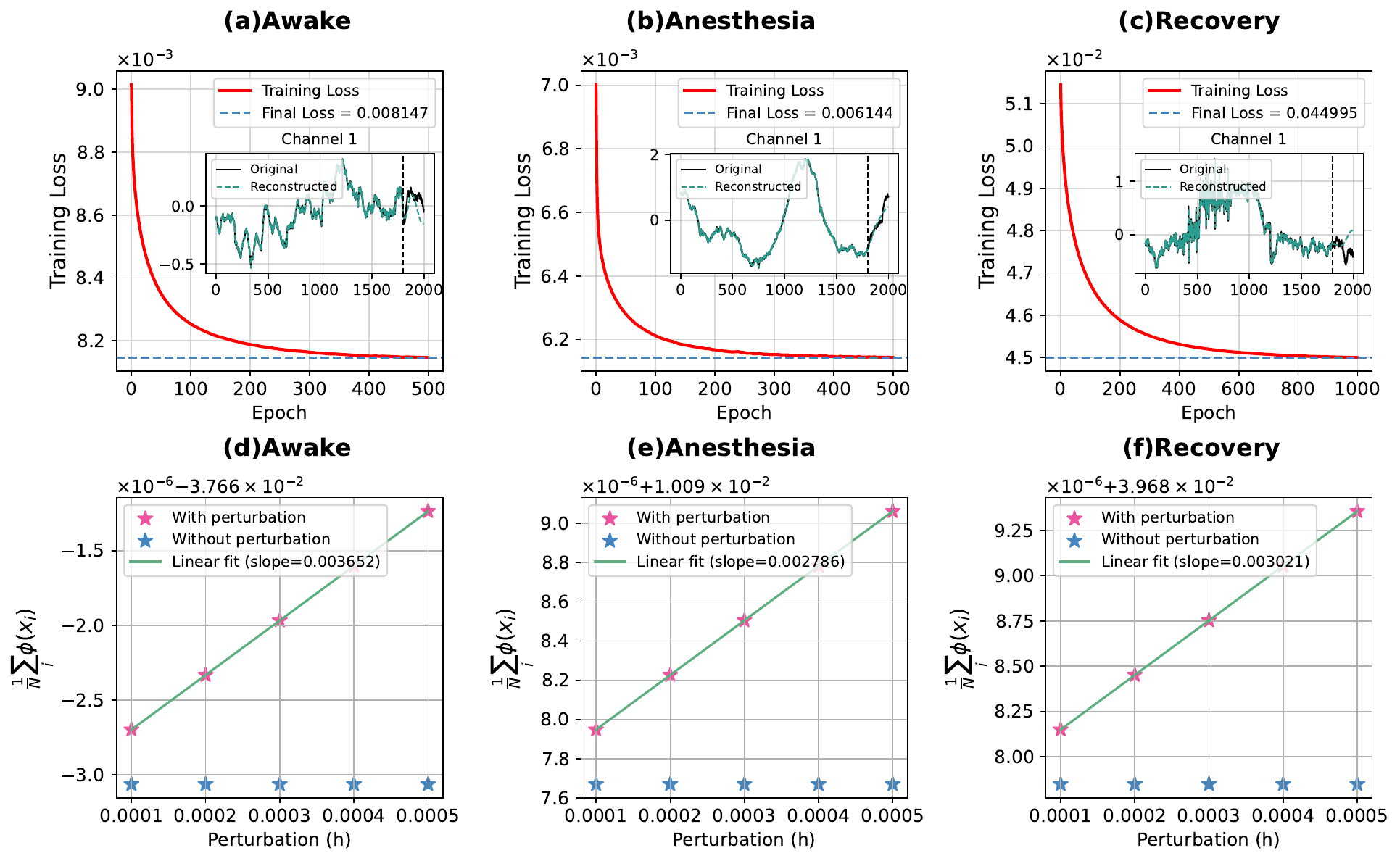}
  \caption{ 
   \blue{Fitted RNN results across three brain states: awake, anesthesia, and recovery.  
    (a-c) Each panel shows the average training loss within $100$-second fitting windows (models were fitted every two seconds).  
    Insets compare the reconstructed and target trajectories. At the $1800$-th time step (indicated by the vertical dashed line), the reconstructed RNN initialized from the target trajectory runs for the next $200$ time steps, in order to evaluate the prediction accuracy.  
    (d-f) The measured mean population activity $\langle \phi(\mathbf{x}(t)) \rangle$ as a function of perturbation amplitude $h$ (applied at $t=0.005$~s, i.e., $5$ time steps away from the moment when the perturbation is applied) for the three different brain states.  The results are average over all inferred recurrent neural networks [see (a-c)].
    The corresponding slope reflects the strength of the neural response function defined in the main text.}
  }
  \label{fig2}
\end{figure}

\begin{figure}[ht]
  \centering
  \includegraphics[width=0.85\textwidth]{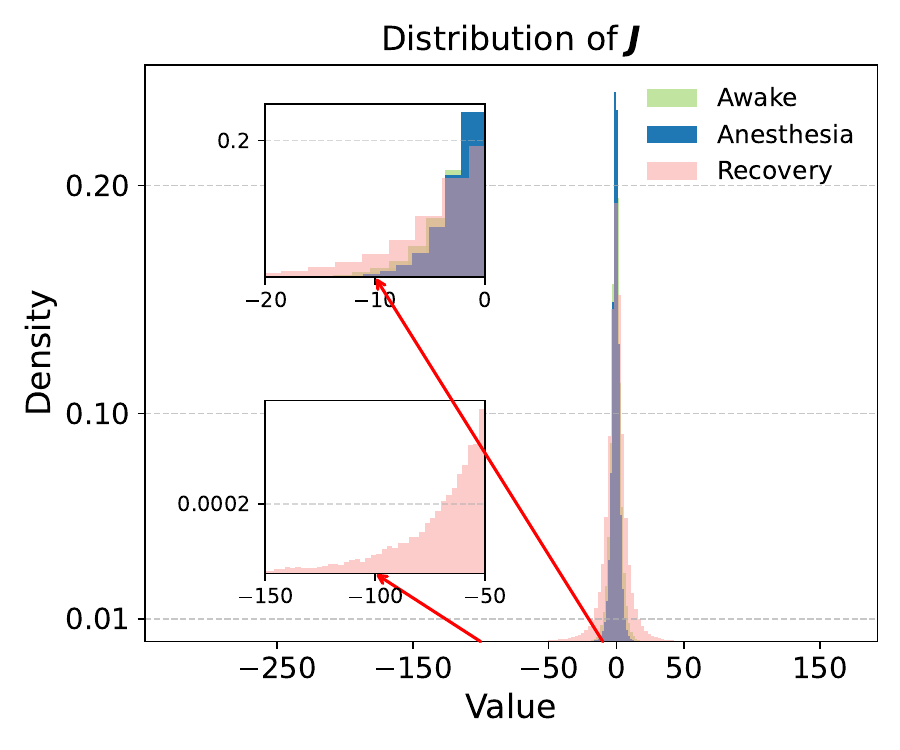}
  \caption{
 \blue{ Distribution of the learned recurrent weight matrix $\J$ across the three brain states: awake, anesthesia, and recovery.  
    All learned coupling weights obtained from $100$-second fitting windows (as in Fig.~\ref{fig2}) are combined and shown in a single plot, with different colors representing different brain states. 
    The inset provides an enlarged view that highlights subtle differences in the weight distributions among the three states. }
  }
  \label{fig3}
\end{figure}

\begin{figure}[ht]
  \centering
  \includegraphics[width=0.95\textwidth]{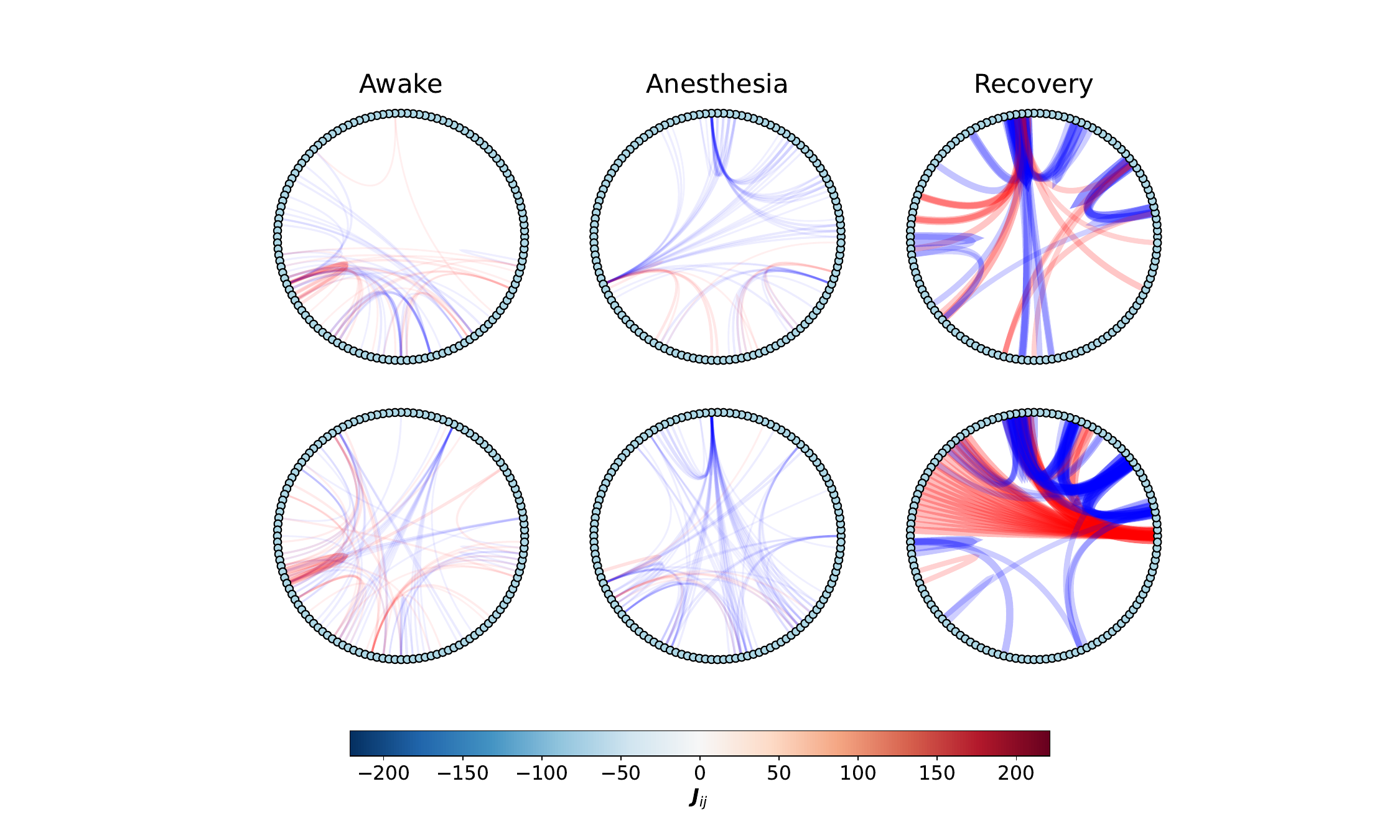}
  \caption{
  \blue{Connection pattern of $\J$ for the three brain states: awake, anesthesia, and recovery.  
    Results are shown for the first two-second segment from the $100$-second fitting window.  
    Shown are the strongest 1\% of connections in absolute value among the $128$ channels.  
    Positive weights are shown in red and negative weights in blue.  
    The first row displays the upper-triangular part of $\J$, and the second row shows the lower-triangular part, highlighting how coupling patterns evolve across different cortical conditions.}
    }
  \label{fig4}
\end{figure}

\section{Results and discussion}
In this section, we show the results of our methodology applied to the Monkey's experimental data, focusing on the transition between conscious and unconscious brain states.

\subsection{RNN model reveals state-dependent network structures}
We first assessed how well the RNN model captures the local dynamics of cortical activity in the three consecutive brain states with different levels of consciousness: awake, anesthesia, and recovery. 
The mean reconstruction loss for each two-second ECoG segment was computed and then averaged over longer time windows ($100$~s each). \blue{As shown in Fig.~\ref{fig2} (a-c),  the RNN learns very well the target time series in the brain. However, if the learned RNN is initialized from the target [e.g., at the $1800$-th time step, see the insets of Fig.~\ref{fig2} (a-c)], the later dynamics deviates from the target, which can be explained by the chaotic nature of the brain dynamics, as also evidenced in a recent theoretical study~\cite{Yu-2025}.
How the fitted RNNs respond to small external perturbations is also shown in Fig.~\ref{fig2}, and we will come to discuss these observed differences later.
}

\begin{figure}[ht]
  \centering
  \includegraphics[width=0.95\textwidth]{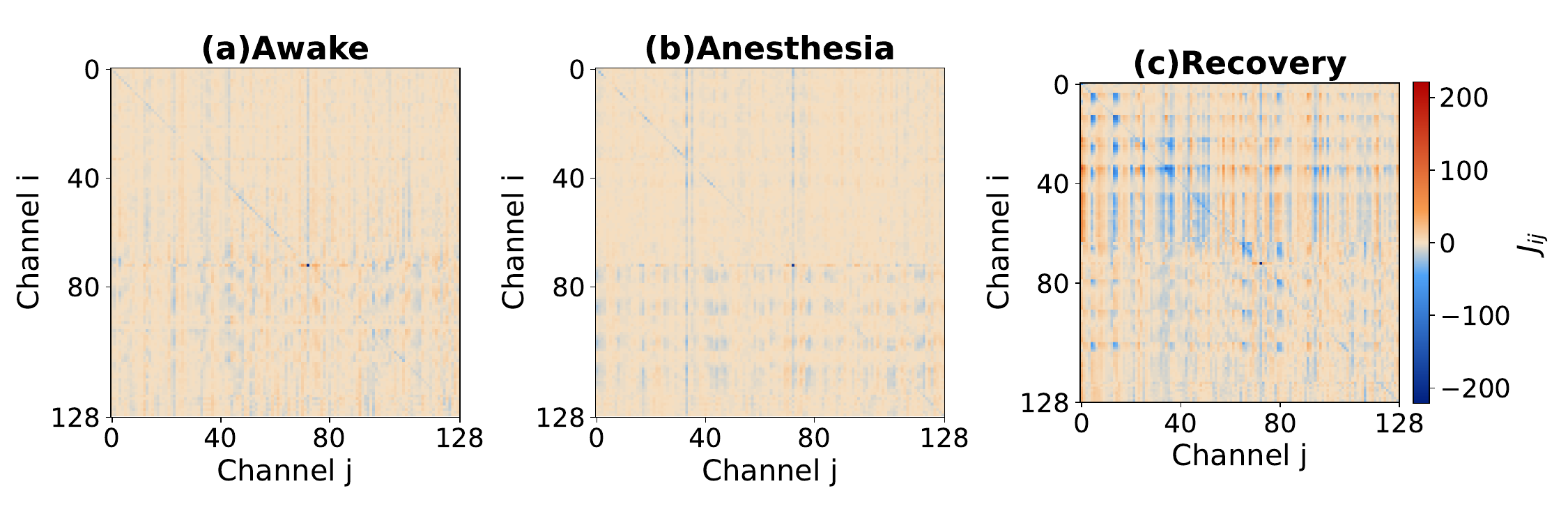}
  \caption{
  \blue{Full recurrent weight matrices $\J$ for the three brain states: awake, anesthesia, and recovery.  
    Each panel shows the entire structured $\J$ matrix (including all elements) corresponding to the same fitted segment as in Fig.~\ref{fig4}, visualized as a heat map.  
    Positive weights are shown in red and negative weights in blue.  }
    }
  \label{fig5}
\end{figure}

 \blue{We next discuss the functional connection patterns learned from the brain dynamics across the three different brain states.}  As shown in Fig.~\ref{fig3}, Fig.~\ref{fig4} and Fig.~\ref{fig5},
 the three stages exhibit clear differences in statistics of functional connectivity. Compared to the awake condition, the anesthesia condition yields sparser connections with weaker strengths, reducing feedback and thus blocking global communications among brain regions~\cite{Massimini-2005}. Nevertheless, the recovery condition yields a broader distribution of coupling, being of salient strength and heterogeneity; these well-tuned feedbacks are able to support consciousness or awareness, showing how consciousness is re-established, which will be further confirmed by the precise connection pattern and the dynamics response property in the next section.

The learned recurrent weight matrices $\mathbf{J}$ are apparently asymmetric, showing consistent differences across three sessions (see Fig.~\ref{fig4} and Fig.~\ref{fig5}). Visualizations of the upper and lower triangles of the $128 \times 128$ coupling matrix reveal that the awake and recovery conditions bear relatively strong and dense but highly heterogeneous couplings, consistent with a rich, flexible, and responsive dynamics (discussed below). During anesthesia, the connection weights become weaker and sparser, indicating a globally less fine-tuned feedback, while the recovery shows an emergence of stronger weights and a structurally-organized connection pattern. For the awake, especially recovered brain state, a structurally well-organized pattern of strong positive and negative couplings supports the high-dimensional dynamics and robust responsiveness. Anesthesia flattens this structure, making connections sparse and suppressing global communication~\cite{Massimini-2005}. Recovery restores this structural heterogeneity, signaling the gradual return of responsiveness as quantified by $\chi(t)$. Results shown in Fig.~\ref{fig4} and Fig.~\ref{fig5} also suggest that in terms of the connection pattern, the recovery state after the loss of consciousness is distinct from the awake state just before the anesthesia, which deserves future systematic analysis.

\begin{figure}[ht]
  \centering
  \includegraphics[width=0.95\textwidth]{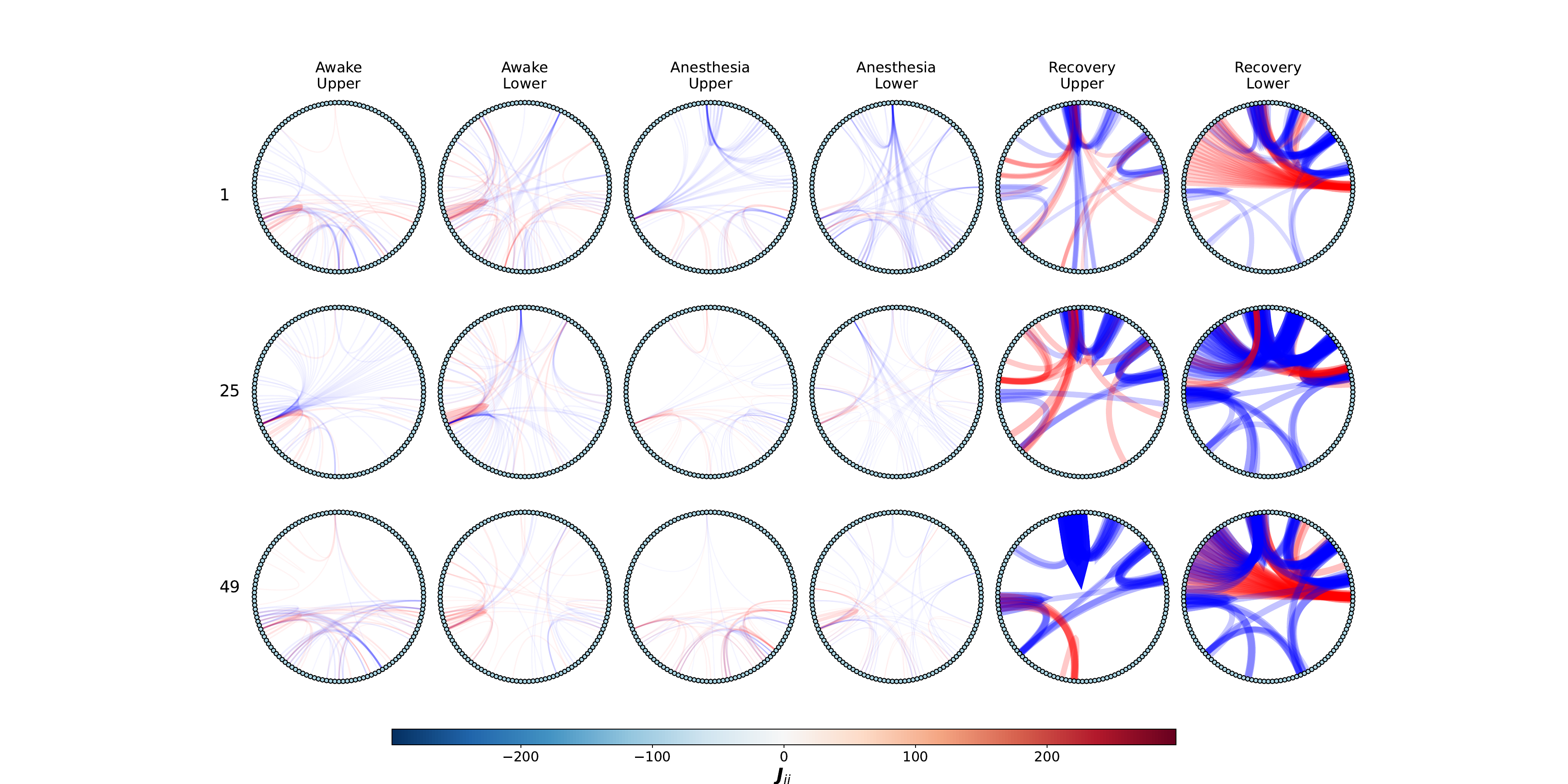}
  \caption{\blue{ Comparison of recurrent connection patterns across the three brain states---awake, anesthesia, and recovery.  
    For each brain state, the plots display the recurrent connections $\J$ obtained from the first, middle, and last two-second segments (indicated by the numbers on the left-hand side) within the $100$-second fitting window (as in Fig.~\ref{fig4}), allowing direct comparison of inferred functional connectivity structures across time.  For each condition, the upper- and lower-triangular parts of the weight matrix are separately displayed.}
  }
  \label{fig6}
\end{figure}

\begin{figure}[ht]
  \centering
  \includegraphics[width=0.95\textwidth]{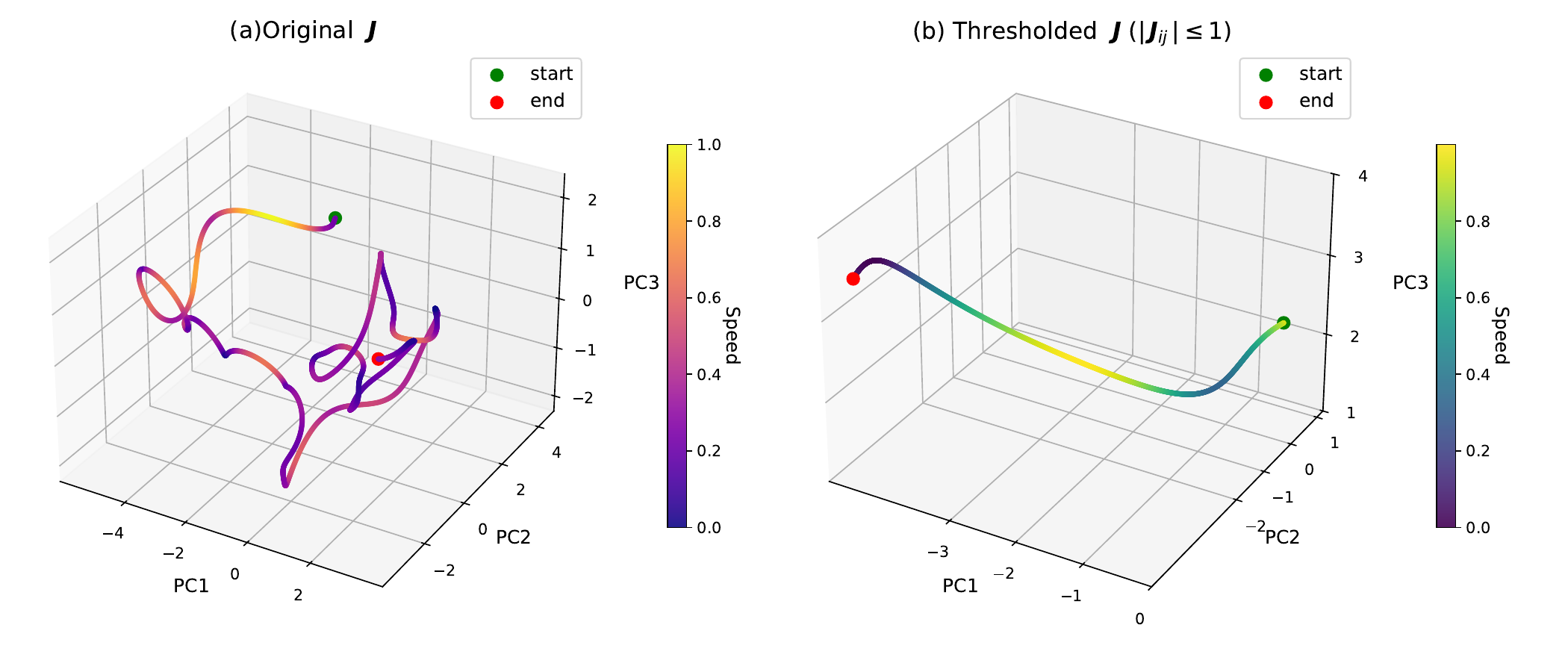}
  \caption{
\blue{Neural trajectories around the edge of instability in recurrent neural networks learned from a two-second segment. The trajectories are colored by their speeds (the magnitude of the velocity $\dot{\bx}$). 
    (a) Trajectories in 3D PCA (principal component analysis) space generated from the original fitted recurrent weights (awake state), showing irregular and high-dimensional patterns consistent with chaotic dynamics.  
    (b) Trajectories after setting all recurrent weights greater than $1$ in absolute value to zero, showing convergence toward a low-dimensional fixed point.  
    This dynamics visualization in both unperturbed and perturbed settings manifests the edge of instability, further supporting our response function angle and the Lyapunov exponent analysis.}
  }
  \label{fig7}
\end{figure}

\blue{ We also verify how the connection pattern varies across different segments from which the functional connectivity is learned. As shown in Fig.~\ref{fig6}, the qualitative property of each connection pattern holds across different
segments within each individual brain state.
To see how these connection patterns impact the dynamics, we also do a control experiment where the original learned couplings for a given inferred network are pruned once their absolute values are above a threshold (set to $1$ here, yet a slight increase of the threshold does not affect the qualitative conclusion).
We then compare the controlled dynamics with the original one with an intact topology. Results are shown in Fig.~\ref{fig7}, which demonstrates that the pruned strong feedbacks through the coupling weights play a vital role in maintaining the 
high-dimensional chaotic attractors. In the absence of these strong feedbacks, the dynamical regime shift to a low-dimensional fixed point, and thus the flexible and collective computational function offered by the high-dimensional brain activity disappear. This observation highlights the important role of criticality even in real brain data analysis, as the concept was previously explored in theoretical and computational frameworks~\cite{EoC-2004,PRX-2022, Aihara-2023,Qiu-2025}. 
}

\subsection{Neural response function quantifies the level of consciousness}
To probe how the fitted RNNs respond to small external perturbations within each conscious state, we computed the mean population response function $\chi(t)$ by applying constant weak
 input perturbations and measuring the resulting relationship between mean activity and the perturbation. For each experimental session, Fig.~\ref{fig2} illustrate this linear regime where the neural response function can be read off. The approximately linear trends in these plots demonstrate that the fitted dynamics yield well-defined response functions, with the slope of each line quantifying the network's global sensitivity to external input, which we identify as a metaphor of responsiveness in brain dynamics.  In particular, the anesthesia condition shows slightly smaller slopes, indicating weaker responsiveness, whereas the awake and recovery conditions produce larger linear coefficients, consistent with more flexible non-equilibrium dynamics. These within-state analyses establish that the RNN models capture a well-defined linear relationship between network response and input perturbations, providing a principled estimate of global dynamical sensitivity as also observed in a toy random model~\cite{Qiu-2025}.

\begin{figure}[ht]
  \centering
  \includegraphics[width=0.8\textwidth]{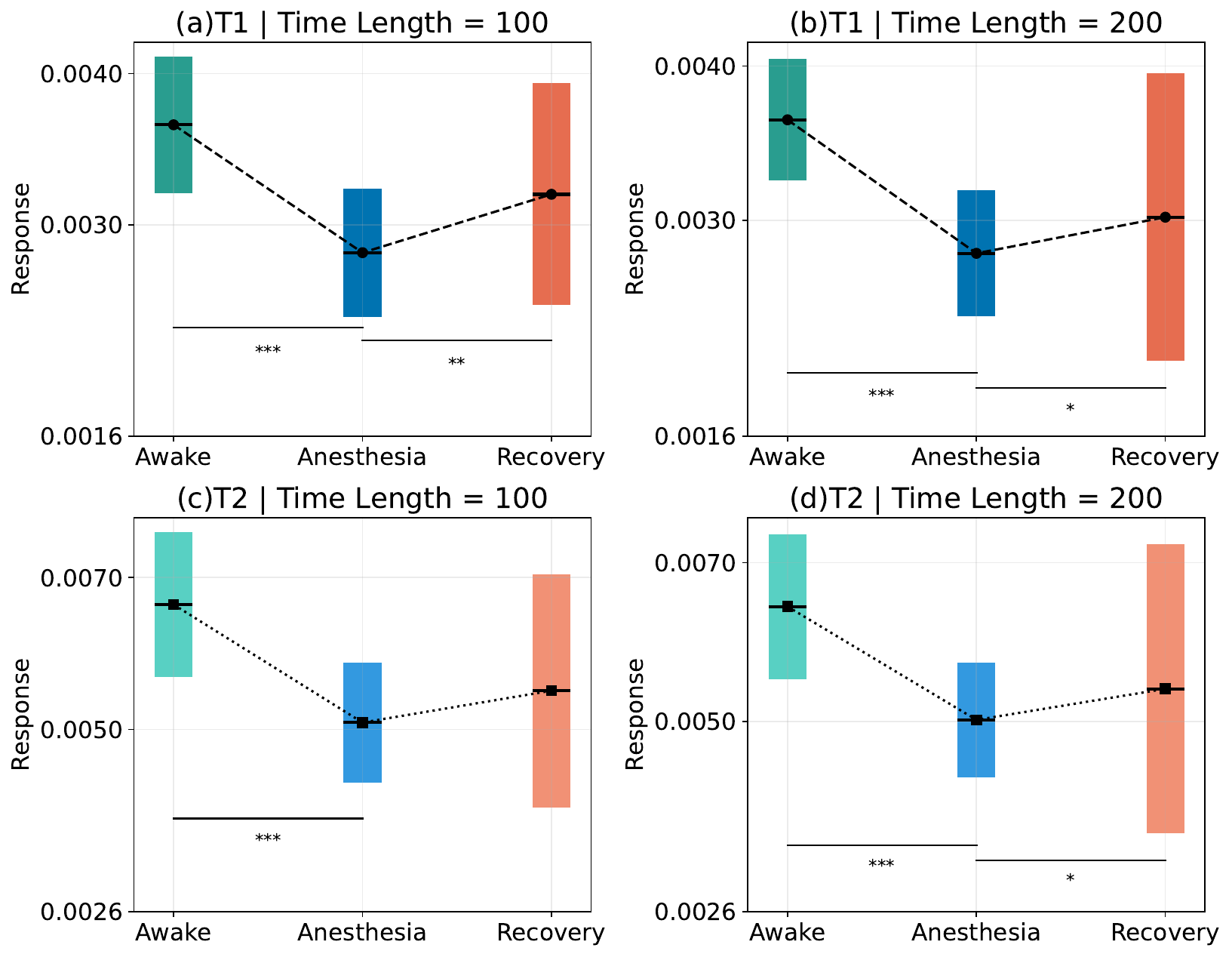}
  \caption{
   \blue{Comparison of neural response functions across brain states: awake, anesthesia, and recovery.  
    (a,b) Average response slopes $\chi(t)$ at $t = 0.005$~s ($5$ time steps, T1 in the plot) for models fitted on all non-overlapping segments of $100$-second (a) and $200$-second (b) windows.  
    (c,d) Average response slopes $\chi(t)$ at $t = 0.01$~s ($10$ time steps, T2 in the plot) for models fitted on all non-overlapping segments of $100$-second (c) and $200$-second (d) windows.  
    Bars represent mean slopes across segments, and error bars indicate standard deviations.  
    Statistical significance between states was assessed with two-sample t-tests (the symbols ``$***$", ``$**$" and ``*" mean the p-value $p < 0.001$, $0.01$, and $0.05$ respectively).
    Note that in the subplot (c), the p-value for the case of anesthesia versus recovery is $0.0938$, and thus there is no symbol shown in the plot.  
    The anesthesia state shows consistently weaker and more concentrated responses, while awake and recovery states show stronger and more variable responsiveness. }
  }
  \label{fig8}
\end{figure}

 To directly compare how this responsiveness varies across conscious states, we next collect these slopes over multiple steady-state time windows and visualize their distributions to verify an indicator of state transition in terms of our 
 theoretically-grounded response function. As shown in Fig.~\ref{fig8}, the anesthesia state consistently shows lower mean responsiveness, highlighting the reduced capacity of the cortex to propagate perturbations under general anesthesia~\cite{Massimini-2005}. In contrast, both the awake and recovery conditions display higher mean response functions, reflecting richer metastable nonequilibrium dynamics that can sustain information processing. Interestingly, the recovery condition just after the loss of consciousness displays a significant variability in neural response function, and moreover, the responsiveness is still weaker than that of the awake condition. \blue{This asymmetric property is interesting and deserves further theoretical studies of underlying mechanisms.}

\begin{figure}[ht]
  \centering
  \includegraphics[width=0.9\textwidth]{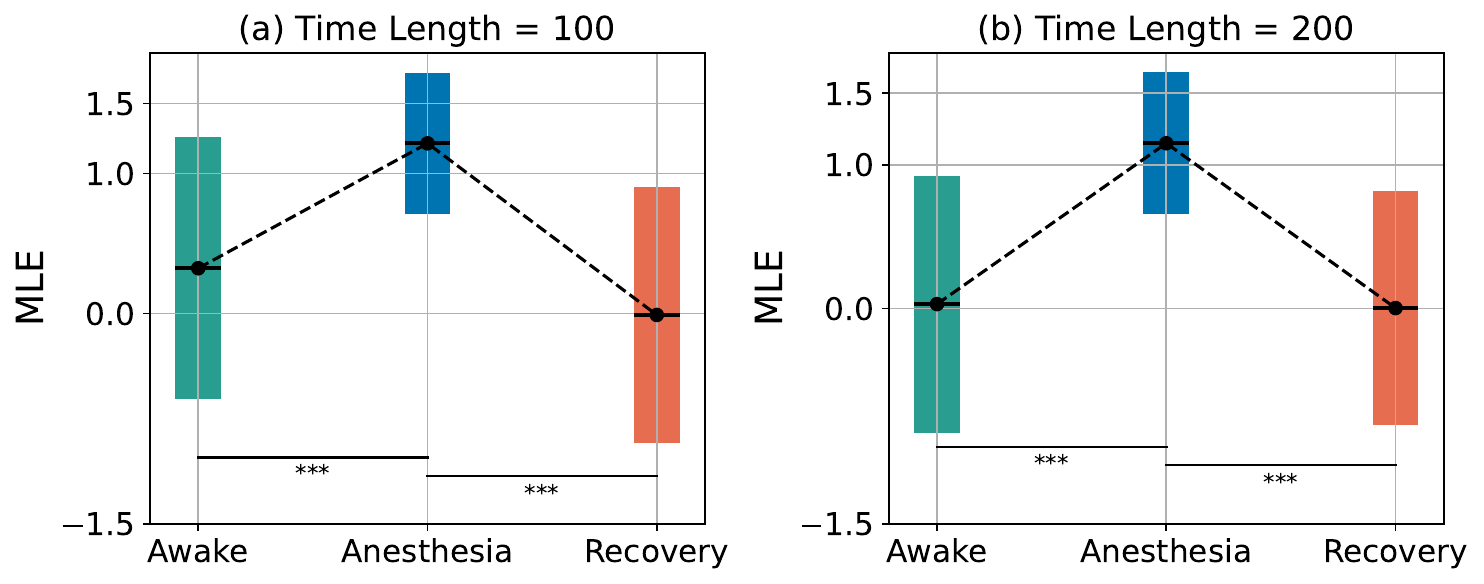}
  \caption{
 \blue{Comparison of dynamic complexity across brain states: awake, anesthesia, and recovery, quantified by the maximal Lyapunov exponent (MLE) using the orbit separation method.  
    (a) MLEs for models fitted on all non-overlapping segments of $100$-second windows.  
    (b) MLEs  for models fitted on all non-overlapping segments of $200$-second windows.  
    The symbols ``$***$" means the p-value $p < 0.001$ for two-sample t-tests.
    The error bars reflect the variability across many fitted segments.  
    The anesthesia state consistently exhibits the highest MLE, indicating greater dynamic instability compared with awake and recovery states.  
    Other settings are the same as in Fig.~\ref{fig8}. }
     }
  \label{fig9}
\end{figure}

Finally, we explore the dynamic complexity of fitted RNNs for three separate sessions in Fig.~\ref{fig9}. The results demonstrate that
the RNNs' dynamics reflect the significant difference across three sessions. More precisely, the unconscious brain state induced by anesthesia bears a higher dynamic instability characterized by maximal Lyapunov exponent (obtained by the orbit separation method~\cite{JC-2003,Yu-2025}), consistent with an empirical study on the relationship between the brain complexity and the dynamic complexity~\cite{PNAS-2022}. Figure~\ref{fig9} also shows the asymmetric role of awake and recovery conditions, consistent with the above connection pattern and response analyses. 

Taken together, these results demonstrate that the fitted RNN framework captures both local dynamical properties (via loss and weight distribution) and global responsiveness (via the response function). The intrinsic differences across states reinforce the view that neural responsiveness is a necessary dynamical signature of consciousness~\cite{PRR-2020, Deco-2019}. By putting these analyses within a nonequilibrium dynamical framework, our work provides a quantitative and mechanistic link between the empirical ECoG signals and theoretical models of cortical criticality and perturbation propagation.
To conclude, these analyses illustrate how local model fitting, recurrent coupling, and perturbation sensitivity collectively reveal state-dependent signatures of consciousness in large-scale cortical dynamics.

\section{Concluding remarks}
How to provide a behavior evaluation of consciousness level is an important frontier, as the quantitative measure helps to characterize brain complexity of consciousness, e.g., clinical treatment of prolonged disorders of consciousness,
unresponsive wakefulness syndrome, and even detection of awareness in understanding the nature of consciousness. Inspired by our previous theoretical works, we argued in this work that the response function of non-equilibrium dynamics in the neural
circuits (here modeled by a recurrent neural network) serves as a natural and simple measurable signature of consciousness, as we observe a salient change of this quantity when analyzing the different stages of ECoG dynamics (e.g,, waking-anesthesia-recovery transition). Another dominant measure in clinical applications is the perturbational complexity index, which provides a direct measure of the spatiotemporal complexity of the evoked responses of the brain to a perturbation such as transcranial magnetic stimulation~\cite{Brain-2020b,PCI-2019}. Computing this index requires multiple steps of delicate analysis (see details in previous works~\cite{PCI-2013,PCI-2019}). It is thus interesting to compare our method with this complexity index in the human dataset. 

Responsiveness studied in this paper is merely a necessary signature of consciousness, while this signature can be mathematically formulated using functionally connected neurons or brain regions. In the future, one has to introduce more biological constraints into the hierarchical brain system, study the multi-scale brain dynamics, and finally demonstrate the sufficient condition to support the maximal information processing ability around a self-organized criticality. The scientific endeavor of putting a data-driven model of brain dynamics within a mathematical framework would be valuable to refine existing theories of consciousness~\cite{CST-2024,CST-2022}, or to provide criteria for justifying or refuting the existence of consciousness in AI systems~\cite{CSAI-2023}.

\section*{Acknowledgments}
This research was supported by the National Natural Science Foundation of China for
Grant number 12475045, and Guangdong Provincial Key Laboratory of Magnetoelectric Physics and Devices (No. 2022B1212010008), and Guangdong Basic and Applied Basic Research Foundation (Grant No. 2023B1515040023).

\section*{Code availability}
Codes to reproduce all results are deposited in our Github~\cite{code-2025}.




\begin{thebibliography}{10}

\bibitem{Neuron-2011}
Stanislas Dehaene and Jean-Pierre Changeux.
\newblock Experimental and theoretical approaches to conscious processing.
\newblock {\em Neuron}, 70(2):200--227, 2011.

\bibitem{TrN-2020}
Tim Bayne, Anil~K. Seth, and Marcello Massimini.
\newblock Are there islands of awareness?
\newblock {\em Trends in Neurosciences}, 43(1):6--16, 2020.

\bibitem{Nature-2025}
Mariana Lenharo.
\newblock The quest to detect consciousness — in all its possible forms.
\newblock {\em Nature}, 643:1172--1174, 2025.

\bibitem{PCI-2013}
Adenauer~G. Casali, Olivia Gosseries, Mario Rosanova, Mélanie Boly, Simone
  Sarasso, Karina~R. Casali, Silvia Casarotto, Marie-Aurélie Bruno, Steven
  Laureys, Giulio Tononi, and Marcello Massimini.
\newblock A theoretically based index of consciousness independent of sensory
  processing and behavior.
\newblock {\em Science Translational Medicine}, 5(198):198ra105--198ra105,
  2013.

\bibitem{JNS-2015}
Guillermo Solovey, Leandro~M Alonso, Toru Yanagawa, Naotaka Fujii, Marcelo~O
  Magnasco, Guillermo~A Cecchi, and Alex Proekt.
\newblock Loss of consciousness is associated with stabilization of cortical
  activity.
\newblock {\em Journal of Neuroscience}, 35(30):10866--10877, 2015.

\bibitem{PCI-2019}
Renzo Comolatti, Andrea Pigorini, Silvia Casarotto, Matteo Fecchio, Guilherme
  Faria, Simone Sarasso, Mario Rosanova, Olivia Gosseries, Mélanie Boly,
  Olivier Bodart, Didier Ledoux, Jean-François Brichant, Lino Nobili, Steven
  Laureys, Giulio Tononi, Marcello Massimini, and Adenauer~G. Casali.
\newblock A fast and general method to empirically estimate the complexity of
  brain responses to transcranial and intracranial stimulations.
\newblock {\em Brain Stimulation}, 12(5):1280--1289, 2019.

\bibitem{Huang-2024}
Haiping Huang.
\newblock Eight challenges in developing theory of intelligence.
\newblock {\em Front. Comput. Neurosci}, 18:1388166, 2024.

\bibitem{PNAS-2021}
Christopher~W. Lynn, Eli~J. Cornblath, Lia Papadopoulos, Maxwell~A. Bertolero,
  and Danielle~S. Bassett.
\newblock Broken detailed balance and entropy production in the human brain.
\newblock {\em Proceedings of the National Academy of Sciences},
  118(47):e2109889118, 2021.

\bibitem{PRE-2021}
Yonatan Sanz~Perl, Hern\'an Bocaccio, Carla Pallavicini, Ignacio
  P\'erez-Ipi\~na, Steven Laureys, Helmut Laufs, Morten Kringelbach, Gustavo
  Deco, and Enzo Tagliazucchi.
\newblock Nonequilibrium brain dynamics as a signature of consciousness.
\newblock {\em Phys. Rev. E}, 104:014411, 2021.

\bibitem{Galad-2021}
Javier~A Galad{\'\i}, S~Silva Pereira, Y~Sanz Perl, Morten~L Kringelbach,
  I~Gayte, Helmut Laufs, Enzo Tagliazucchi, Jos{\'e}~A Langa, and Gustavo Deco.
\newblock Capturing the non-stationarity of whole-brain dynamics underlying
  human brain states.
\newblock {\em NeuroImage}, 244:118551, 2021.

\bibitem{PRE-2023}
Matthieu Gilson, Enzo Tagliazucchi, and Rodrigo Cofr\'e.
\newblock Entropy production of multivariate ornstein-uhlenbeck processes
  correlates with consciousness levels in the human brain.
\newblock {\em Phys. Rev. E}, 107:024121, 2023.

\bibitem{BD-2025}
Ramón Nartallo-Kaluarachchi, Morten~L. Kringelbach, Gustavo Deco, Renaud
  Lambiotte, and Alain Goriely.
\newblock Nonequilibrium physics of brain dynamics.
\newblock {\em arXiv:2504.12188}, 2025.

\bibitem{FNC-2014}
Leandro~M. Alonso, Alex Proekt, Theodore~H. Schwartz, Kane~O. Pryor,
  Guillermo~A. Cecchi, and Marcelo~O. Magnasco.
\newblock Dynamical criticality during induction of anesthesia in human ecog
  recordings.
\newblock {\em Frontiers in Neural Circuits}, 8, 2014.

\bibitem{PNAS-2022}
Daniel Toker, Ioannis Pappas, Janna~D Lendner, Joel Frohlich, Diego~M Mateos,
  Suresh Muthukumaraswamy, Robin Carhart-Harris, Michelle Paff, Paul~M Vespa,
  Martin~M Monti, et~al.
\newblock Consciousness is supported by near-critical slow cortical
  electrodynamics.
\newblock {\em Proceedings of the National Academy of Sciences},
  119(7):e2024455119, 2022.

\bibitem{Eagleman-2019}
Sarah~L. Eagleman, Divya Chander, Christina Reynolds, Nicholas~T. Ouellette,
  and M.~Bruce MacIver.
\newblock Nonlinear dynamics captures brain states at different levels of
  consciousness in patients anesthetized with propofol.
\newblock {\em PLOS ONE}, 14(10):1--26, 10 2019.

\bibitem{CS-2022}
Laura~Alethia de~la Fuente, Federico Zamberlan, Hernán Bocaccio, Morten
  Kringelbach, Gustavo Deco, Yonatan~Sanz Perl, Carla Pallavicini, and Enzo
  Tagliazucchi.
\newblock Temporal irreversibility of neural dynamics as a signature of
  consciousness.
\newblock {\em Cerebral Cortex}, 33(5):1856--1865, 2022.

\bibitem{Cruzat-2023}
Josephine Cruzat, Ruben Herzog, Pavel Prado, Yonatan Sanz-Perl, Raul
  Gonzalez-Gomez, Sebastian Moguilner, Morten~L. Kringelbach, Gustavo Deco,
  Enzo Tagliazucchi, and Agust{\'\i}n Iba{\~n}ez.
\newblock Temporal irreversibility of large-scale brain dynamics in
  alzheimer{\textquoteright}s disease.
\newblock {\em Journal of Neuroscience}, 43(9):1643--1656, 2023.

\bibitem{Qiu-2025}
Junbin Qiu and Haiping Huang.
\newblock An optimization-based equilibrium measure describing fixed points of
  non-equilibrium dynamics: application to the edge of chaos.
\newblock {\em Communications in Theoretical Physics}, 77(3):035601, 2025.

\bibitem{Du-2024}
Wenkang Du and Haiping Huang.
\newblock Synaptic plasticity alters the nature of chaos transition in neural
  networks.
\newblock {\em arXiv:2412.15592}, 2024.

\bibitem{Eric-2024}
Dana Mastrovito, Yuhan~Helena Liu, Lukasz Kusmierz, Eric Shea-Brown, Christof
  Koch, and Stefan Mihalas.
\newblock Transition to chaos separates learning regimes and relates to measure
  of consciousness in recurrent neural networks.
\newblock {\em bioRxiv}, 2024.

\bibitem{Data-2013}
Toru Yanagawa, Zenas~C. Chao, Naomi Hasegawa, and Naotaka Fujii.
\newblock Large-scale information flow in conscious and unconscious states: an
  ecog study in monkeys.
\newblock {\em PLOS ONE}, 8(11), 11 2013.

\bibitem{Yu-2025}
Zhendong Yu and Haiping Huang.
\newblock Network reconstruction may not mean dynamics prediction.
\newblock {\em Physical Review E}, 111(3):034308, 2025.

\bibitem{Massimini-2005}
Marcello Massimini, Fabio Ferrarelli, Reto Huber, Steve~K. Esser, Harpreet
  Singh, and Giulio Tononi.
\newblock Breakdown of cortical effective connectivity during sleep.
\newblock {\em Science}, 309(5744):2228--2232, 2005.

\bibitem{tycho}
https://neurotycho.org/.

\bibitem{Chaos-1988}
H.~Sompolinsky, A.~Crisanti, and H.~J. Sommers.
\newblock Chaos in random neural networks.
\newblock {\em Phys. Rev. Lett.}, 61:259--262, 1988.

\bibitem{FD-2008}
Umberto Marini~Bettolo Marconi, Andrea Puglisi, Lamberto Rondoni, and Angelo
  Vulpiani.
\newblock Fluctuation–dissipation: Response theory in statistical physics.
\newblock {\em Physics Reports}, 461(4):111--195, 2008.

\bibitem{Zou-2024}
Wenxuan Zou and Haiping Huang.
\newblock {Introduction to dynamical mean-field theory of randomly connected
  neural networks with bidirectionally correlated couplings}.
\newblock {\em SciPost Phys. Lect. Notes}, page~79, 2024.

\bibitem{EoC-2004}
Nils Bertschinger and Thomas Natschl{\"a}ger.
\newblock Real-time computation at the edge of chaos in recurrent neural
  networks.
\newblock {\em Neural computation}, 16(7):1413--1436, 2004.

\bibitem{PRX-2022}
Kamesh Krishnamurthy, Tankut Can, and David~J. Schwab.
\newblock Theory of gating in recurrent neural networks.
\newblock {\em Phys. Rev. X}, 12:011011, Jan 2022.

\bibitem{Aihara-2023}
Takashi Kanamaru, Takao~K. Hensch, and Kazuyuki Aihara.
\newblock Maximal memory capacity near the edge of chaos in balanced cortical
  e-i networks.
\newblock {\em Neural Computation}, 35(8):1430--1462, 2023.

\bibitem{JC-2003}
Julien~Clinton Sprott.
\newblock {\em {Chaos and Time-Series Analysis}}.
\newblock Oxford University Press, 2003.

\bibitem{PRR-2020}
A.~Sarracino, O.~Arviv, O.~Shriki, and L.~de~Arcangelis.
\newblock Predicting brain evoked response to external stimuli from temporal
  correlations of spontaneous activity.
\newblock {\em Phys. Rev. Res.}, 2:033355, Sep 2020.

\bibitem{Deco-2019}
Gustavo Deco, Josephine Cruzat, Joana Cabral, Enzo Tagliazucchi, Helmut Laufs,
  Nikos~K. Logothetis, and Morten~L. Kringelbach.
\newblock Awakening: Predicting external stimulation to force transitions
  between different brain states.
\newblock {\em Proceedings of the National Academy of Sciences},
  116(36):18088--18097, 2019.

\bibitem{Brain-2020b}
Dmitry~O. Sinitsyn, Alexandra~G. Poydasheva, Ilya~S. Bakulin, Liudmila~A.
  Legostaeva, Elizaveta~G. Iazeva, Dmitry~V. Sergeev, Anastasia~N. Sergeeva,
  Elena~I. Kremneva, Sofya~N. Morozova, Dmitry~Yu. Lagoda, Silvia Casarotto,
  Angela Comanducci, Yulia~V. Ryabinkina, Natalia~A. Suponeva, and Michael~A.
  Piradov.
\newblock Detecting the potential for consciousness in unresponsive patients
  using the perturbational complexity index.
\newblock {\em Brain Sciences}, 10(12), 2020.

\bibitem{CST-2024}
Johan~F. Storm, P.~Christiaan Klink, Jaan Aru, Walter Senn, Rainer Goebel,
  Andrea Pigorini, Pietro Avanzini, Wim Vanduffel, Pieter~R. Roelfsema,
  Marcello Massimini, Matthew~E. Larkum, and Cyriel M.~A. Pennartz.
\newblock An integrative, multiscale view on neural theories of consciousness.
\newblock {\em Neuron}, 112(10):1531--1552, 2024.

\bibitem{CST-2022}
Anil~K. Seth and Tim Bayne.
\newblock Theories of consciousness.
\newblock {\em Nature Reviews Neuroscience}, 23(7):439--452, 2022.

\bibitem{CSAI-2023}
Patrick Butlin, Robert Long, Eric Elmoznino, Yoshua Bengio, Jonathan Birch,
  Axel Constant, George Deane, Stephen~M. Fleming, Chris Frith, Xu~Ji, Ryota
  Kanai, Colin Klein, Grace Lindsay, Matthias Michel, Liad Mudrik, Megan A.~K.
  Peters, Eric Schwitzgebel, Jonathan Simon, and Rufin VanRullen.
\newblock Consciousness in artificial intelligence: Insights from the science
  of consciousness.
\newblock {\em arXiv:2308.08708}, 2023.

\bibitem{code-2025}
Wenkang Du.
\newblock https://github.com/Wenkang-Du/Response-function-of-consciousness,
  2025.

\end{thebibliography}

\end{document}